\documentclass[twocolumn]{osa-article}
\journal{osajournal}

\articletype{Research Article}

\begin{document}

\title{Optimization for maximum modulation of a double-pass twisted nematic liquid crystal display}

\author{Sebasti\'an Bordakevich,\authormark{1,2} Lorena Reb\'on,\authormark{3,4} and Silvia Ledesma\authormark{1,2,*}}

\address{\authormark{1}Departamento de F\'isica, FCEyN, Universidad de Buenos Aires, Pabell\'on 1, Ciudad Universitaria, Buenos Aires (1428), Argentina\\
\authormark{2}Consejo Nacional de Investigaciones Cient\'ificas y T\'ecnicas (CONICET), Argentina\\
\authormark{3}Departamento de F\'isica, FCE, Universidad Nacional de La Plata, C.C. 67, La Plata (1900), Argentina\\
\authormark{4}Instituto de F\'isica de La Plata, UNLP - CONICET, Argentina}

\email{\authormark{*}ledesma@df.uba.ar} 


\begin{abstract*}
Spatial light modulators are widely used to perform modulations of different properties of the electromagnetic field. In this work, a simple optimization method for general double-pass setups was developed. It takes into account the involved polarizing elements and displays, and a numerical simulation based on an exhaustive search routine finds the optimal optical axes orientation of the polarizing elements for the desired modulation. By simultaneously considering both impingements, we are able to take full advantage of the modulation capabilities of the chosen spatial light modulators. In particular, different polarization modulations and complex amplitude modulations were studied for twisted nematic liquid crystal displays and passive linear optical elements. Examples for different optimization criteria are shown and compared with experimental results, supporting the feasibility of this approach. This method offers the possibility of independent modulation of two properties of the input light state, outperforming the use of a single screen.
\end{abstract*}


\section{Introduction}
Liquid crystal displays (LCDs) can be used as spatial light modulators (SLMs), since they are programmable structures capable of modifying, pixel by pixel, an incoming electromagnetic field and, in combination with passive linear optics, they can modulate properties such as intensity, phase or polarization. SLMs are widely used in several applications \cite{maurer2010,trichili2016,peinado2010,rubinsztein2017,bhebhe2018,flamini2018,pabon2019,lohrmann2019}, including microscopy, optical communications, polarimetry, structured light, optical tweezers and quantum information processing. For many of those implementations it is required to have a wide range of modulation, with high precision, in one or more properties of the incoming beam. However, such required modulation can be difficult to achieve due to different experimental reasons. In particular, a considerable variability in the modulation capabilities is observed when different LCDs are compared, since they highly depend on the particular technology with which the display is manufactured. This fact entails the need to study each LCD individually and, consequently, find efficient methods to optimize its performance as an SLM, take advantage of its properties, and maximize the desired modulations.\\

Among LCDs, twisted nematic displays stand out for having an excellent response regarding parameters like resolution, refresh rate and filling factor. Furthermore, they are easily available and relatively inexpensive, as they are commonly built to modulate intensity in commercial applications, like videoprojectors. In such displays, the alignment axes of the liquid crystal molecules are different for every layer of the LCD and change with the set gray level \cite{chandrasekhar}. As a consequence, it is not straightforward to properly understand how the properties of light are effectively modified according to the incident polarization and the gray level of the screen. Hence, a variety of characterization techniques has been reported for this purpose \cite{neto1996,marquez2001,kruger2015,chandra2020,tiwari2021}. Moreover, it has been observed that the modulations of the different properties of the electromagnetic field are coupled to each other. This has led to great efforts to decouple them and achieve independent modulations, requiring rigorous characterizations of the displays. In particular, optimizations for intensity-only and phase-only modulations were achieved by selecting the appropriate input and output polarization states. For example, it has been done via a modeling of the LCDs using the Jones formalism and obtaining the parameters of interest for the given screen through transmission measurements \cite{marquez2001}, or studying their polarimetric properties and Mueller matrices by experimentally measuring intensity and phase \cite{marquez2008}, or applying the polar decomposition and combining both frameworks \cite{moreno2008}. As shown in those works, optimizations are crucial to determine an experimental configuration that maximally exploits the features of the screens and, in the best case, allows the system to achieve the desired modulations.\\

However, even when decoupling is accomplished, the problem of a complete modulation still remains. For instance, commercial twisted nematic screens are getting thinner and because phase modulation depends on the thickness of the liquid crystal, complete phase modulation is often not available. As a solution, it has been studied the possibility of using two consecutive twisted nematic LCDs to improve phase modulation \cite{kelly1999,hsieh2007}. Also, complex amplitude modulation is achieved by independently optimizing two displays, the first one working in a phase-only modulation mode, and the second one in an amplitude-only modulation mode\cite{neto1996}, or vice versa \cite{lima2011}. But since the optimal configuration for each display is not necessarily the optimal configuration when both displays are simultaneously considered, this approach does not completely exploit the joint system capabilities. Alternatively, the combination of several pixels into a single macro-pixel to achieve complete complex amplitude modulation was proposed and implemented \cite{vanputten2008,maluenda2013,hasegawa2019}. Furthermore, an efficient codification of the complex amplitude information into phase-only spatial light modulators was developed \cite{davis1999,arrizon2007,bolduc2013,varga2014}. Similar techniques were also applied to generate vector beams \cite{wang2007,rong2014,guo2014}. Nevertheless, these methods significantly reduce the effective resolution of the SLM and require a subsequent spatial filtering to obtain the desired light state. Recently, a double-pass setup was jointly optimized to achieve arbitrary complex modulation, after the light impinges on the two sides of a single twisted nematic LCD between fixed polarizers \cite{macfaden2017}. In that work, the modulation capabilities of the SLMs are described using the Jones matrix formalism which works for fully polarized light.\\

In the present work we consider a double-pass configuration with fixed optics, for which pixels are not combined into macro-pixels, nor any codification is implemented. This allows to make use of the full resolution of the displays while avoiding the need to perform a subsequent decodification. As optical elements, we include waveplates and polarizers, in order to be able to choose arbitrary elliptical input and output polarizations, thus maximizing the amount of possible experimental configurations. Additionally, our approach for optimizing the experimental configuration is based, on the one hand, on Mueller formalism to describe the different optical elements involved, including the SLMs. This formalism allows us to account for the effects of depolarization, which are ignored in the Jones formalism. On the other hand, we also include Jones formalism to determine the additional phase shifts introduced by each of the elements. Following these considerations, a simple and general optimization method was developed with the purpose of performing several modulations of practical interest. Our optimization algorithm can concurrently take into account the effects of two individual displays as well as two consecutive impingements onto the same display, which efficiently exploits the modulation capabilities of the joint system. This allows to obtain the best performance that the available SLMs can offer for any chosen modulation, i.e., to maximize both the dynamic range modulation and the resolution. To this aim, an exhaustive search routine was programmed, from which we can find the optimal experimental setup configuration for several target modulations. Finally, the results of the numerical optimization for two twisted nematic displays, extracted from a commercial video projector, are shown and compared with the corresponding modulations obtained experimentally. They reveal not only a better performance than in the case where a single screen is used, but also the possibility to perform two-dimensional modulations, in which two parameters can be independently modulated. Specifically, we focused on complex amplitude and polarization modulations. At our knowledge, this approach to optimizing polarization modulation is reported here for the first time.

\section{Mathematical description}
The use of two displays is usually required to achieve such two-dimensional modulations, like complex amplitude or polarization, given that the effects of an LCD acting as an SLM are generally described as a function of a single parameter: the gray level. Thus, the properties of the light field after impinging onto the LCD are also dependent on that single variable, which becomes an important constraint when attempting simultaneous arbitrary modulation in two or more of its properties. For example, if the modulation of the complex amplitude of the electromagnetic field is studied, sweeping over the gray level of the screen results in states that follow a curve on the complex plane \cite{juday1991}. Similarly, if the polarization modulation is studied, changing the gray level gives a set of states that describe a curve in the Poincaré sphere \cite{sit2017}. When modulating twice, the resulting states no longer depend on a single variable, thus forming surfaces in those spaces, not curves, and removing that limitation \cite{juday1991, kenny2012}.\\

\begin{figure}[h]
\centering\includegraphics[width=0.9\linewidth]{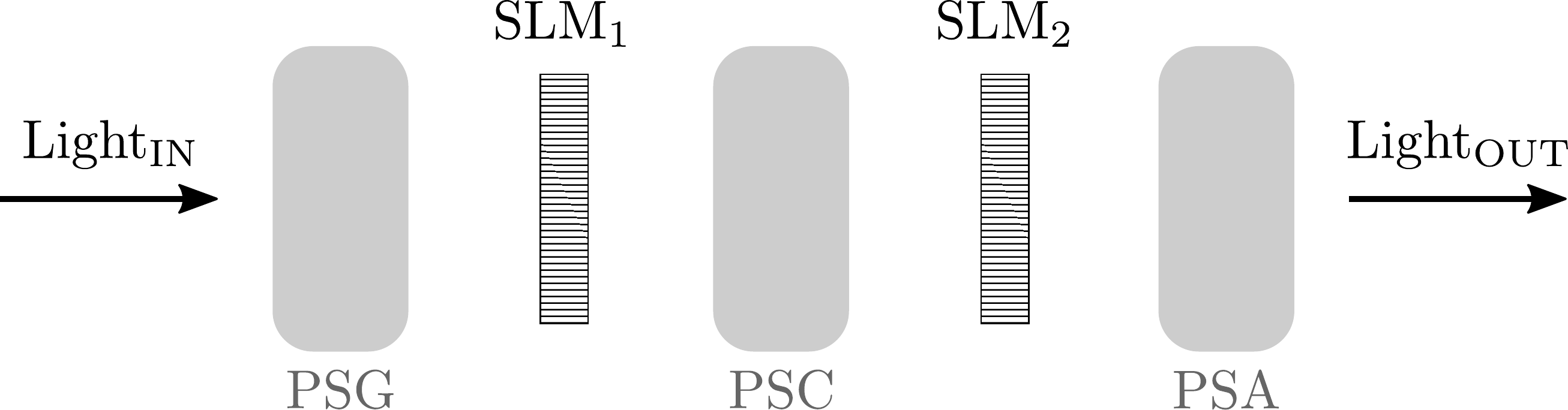}
\caption{Schematic of the optical system under study. It consists of two spatial light modulators (SLM$_i$) in addition to a polarization state generator (PSG), a polarization state coupler (PSC) and a polarization state analyzer (PSA), which comprise passive linear optical elements. Light$_\text{IN}$ and Light$_{\text{OUT}}$ refer to the input and output light states. Each SLM$_i$ may be different zones or sides of the same device.}
\label{fig:setup_generico}
\end{figure}

In particular, the double-pass SLM system that we have studied is schematically shown in Fig. \ref{fig:setup_generico}: it consists of an input light field (Light$_{\text{IN}}$) that is modified by a polarization state generator (PSG), determining the state that reaches the first spatial light modulator (SLM$_1$). The resulting state is modified by a polarization state coupler (PSC) and sent to a second spatial light modulator (SLM$_2$). Finally, the emerging field goes through a polarization state analyzer (PSA), resulting in an output light (Light$_{\text{OUT}}$). Each of these systems consists of passive linear optical elements like waveplates, polarizers or mirrors, and its presence depends on the chosen experimental configuration. For a more general analysis, they are all included in the formalism. On the other hand, the SLMs can be any device with capabilities of modifying the light state according to the selected spatial region of the wavefront, like LCDs, micromirror devices or deformable mirrors. It is worth noting that SLM$_1$ and SLM$_2$ are not necessarily two different devices, but can also be different sides or spatial zones on the screen of the same device.\\

A mathematical description of the physical properties of this system can be made through Mueller calculus \cite{goldstein}. Under this framework, light states are described by the four real parameters of a Stokes vector $S$, and each of the optical elements or systems by a 4-by-4 real matrix. Also, it allows to consider light that may not be completely polarized, and depolarizing optics. The Mueller matrix of the whole system, $M$, operates over the input state $S_{\text{IN}}$, resulting in an output state $S_{\text{OUT}}$.

\begin{eqnarray}\label{eq:mueller}
    S_{\text{OUT}}&=&M\,S_{\text{IN}}\\
    &=&M_{\text{PSA}}\, M_{\text{SLM}_2}\, M_{\text{PSC}}\, M_{\text{SLM}_1}\, M_{\text{PSG}}\,  S_{\text{IN}},\nonumber
\end{eqnarray}

where $M$ was decomposed as the standard product of the matrices of the optical systems in the array.  Additionally, every element may shift the phase of the state, which is not taken into account by Mueller formalism. This fact can be analyzed, separately, with Jones calculus \cite{goldstein}, that describes the field state of the light by a two-dimensional complex vector and the optical elements by 2-by-2 complex matrices. Analogously to Eq.~(\ref{eq:mueller}), the Jones matrix of the system, $J$, operates over the input state $\Bar{E}_{\text{IN}}$, resulting in an output state $\Bar{E}_{\text{OUT}}$: 

\begin{eqnarray}\label{eq:jones}
    \Bar{E}_{\text{OUT}}&=&J\,\Bar{E}_{\text{IN}}\\
    &=&J_{\text{PSA}}\, J_{\text{SLM}_2}\, J_{\text{PSC}}\, J_{\text{SLM}_1}\, J_{\text{PSG}}\,  \Bar{E}_{\text{IN}}.\nonumber
\end{eqnarray}

In this formalism, $\Bar{E}_{\text{OUT}}$ can be defined as  

\begin{equation}\label{eq:jonesvector}
\Bar{E}_{\text{OUT}}=\text{exp}\left(i\beta_1+i\beta_2\right)\begin{pmatrix}|E_x|\,\text{exp}(i\varphi_x)\\ |E_y|\,\text{exp}(i\varphi_y)\end{pmatrix},
\end{equation}

where the phases $\varphi_x$ and $\varphi_y$ depend on the incident polarization, as well as parameters of the displays and the involved optics, and they are predicted by Jones calculus. Additionally, every element provides a global phase shift due to their thickness, which does not depend on the polarization state of light. Given that these do not contribute to any modulation, they are not considered, except for the global phases of the displays, $\beta_1$ and $\beta_2$. These do contribute to the phase modulation due to their dependence on the gray level of each screen, and have to be experimentally determined in relation to a common reference. They are originated in a variation of the optical path length due to the orientation of the liquid crystal molecules.\\

It is important to note that, as the SLMs add spatial dependency to the input light state by individually assigning a gray level to each pixel of the displays, the resulting state can be considered as an image, either in intensity, phase, polarization, or a combination of them. Hence, if an optical element is modified (e.g. the direction of the optical axis of a polarizer or a waveplate is changed) every value in the image gets collectively modified. This changes the range of the accessible outcomes and discards the possibility of independently representing different states in different regions of the display at the same time. So, in order to represent images, the optical elements must remain fixed and the modulations be implemented by only changing the gray levels of the SLMs. This leads to the conclusion that for each particular SLM device, the ability of creating good quality images depends on the possibility of finding the optimal configuration of the given optical elements of the setup.

\section{Computational optimization}
In order to find the best optical configuration for some given modulations, a simulation based on an exhaustive search routine was programmed in \mbox{MATLAB}. We have considered that PSG, PSC and PSA consist of optical elements like mirrors, polarizers whose optical axes can be set to angles $\theta_\text{P}$, or retarders like quarter waveplates whose fast axis can be set to angles $\theta_\text{QW}$. Furthermore, the gray levels of SLM$_1$ and SLM$_2$ were set independently. The routine uses Mueller calculus to predict the polarization state of light obtained by every given combination of PSG, PSC, PSA and gray levels, and Jones calculus for their corresponding phases. Then, by sweeping over those set of gray levels, it results in the collection of states that can be generated for every setting of the optical elements.\\

We present here two of the examples that were optimized for two-parameter modulation. They are (1) \textit{polarization modulation} and (2) \textit{complex amplitude modulation}, but any other desired modulation could be approached in a similar way.

\subsection{Polarization modulation}

To achieve polarization modulation, it is required that the PSA does not have a polarizer in it, otherwise the output electromagnetic field will always arrive to the detector with the same polarization state, despite the change in the gray levels of SLM$_1$ and SLM$_2$. After sweeping over the gray levels and collecting the resulting polarization states for every optical setup, they were placed on the Poincaré sphere according to their Stokes parameters. Then, to find if there is a configuration that allowed us to obtain complete polarization modulation, we have searched for the Poincaré sphere where the mean distance between neighbor states is maximized, for a fixed number of states given by the whole sweeping in the gray levels of SLM$_1$ and SLM$_2$. This leads us to the configuration that allows obtaining polarization states distributed over the largest possible surface of the sphere. In cases where no configuration was found that allows complete modulation, we have considered the optimal configuration to be the one that let us to have access to a larger amount of experimentally distinguishable states. To quantify this separation between states the Euclidean distance in the Poincaré sphere of radius 1 was calculated as \mbox{$d(S^A,S^B)=\sqrt{\sum_{i=0}^3\left(S_i^B-S_i^A\right)^2}$}.\\

In addition, a search for specific polarization states was performed: configurations were selected to include certain states of interest, like linear or circular polarizations. This was done by checking which configurations had at least one outgoing state within a small distance to those target states in the Poincaré sphere.

\subsection{Complex amplitude modulation}

In this case, the intensity of the input electromagnetic field is modulated only after projecting the resulting polarized light onto a certain polarization state. Thus, the presence of a polarizer in the PSA is required, otherwise when changing the gray levels of SLM$_1$ and SLM$_2$, the polarization state of the light will change but not its intensity. After sweeping over the gray levels and collecting the resulting states for every optical setup, they were placed on the complex plane according to their intensity and phase values. Then, we have searched for the complex plane where the displayed states spanned the maximum surface. This was achieved by comparing the areas enclosed by the concave hull that contains the set of states, and selecting the largest ones. To compute the areas, the \texttt{boundary.m} function \cite{boundary} was implemented, setting the shrinking factor to 1. This is equivalent to obtaining the boundary of the $\alpha$-shape with the minimum radius, so that the surface is not split. If  no configuration was found that allows complete modulation, we have considered the optimal configuration to be the one that let us to generate more experimentally distinguishable states, that is, the one where the obtained states spanned the maximum surface. Hence, we have performed a search under specific constraints, for example, configurations that simultaneously include an intensity-only and a phase-only modulation. This was accomplished by choosing those configurations that cover a wide range of phase modulation, while achieving a high value of maximum intensity and a low value of minimum intensity. This was quantified as the contrast, defined as the ratio between the difference and the sum of the maximum and minimum intensity: $\left(\text{I}_\text{max}-\text{I}_\text{min}\right)/\left(\text{I}_\text{max}+\text{I}_\text{min}\right)$.\\

In addition, we searched for configurations that allowed arbitrary complex amplitude modulation if intensity is limited to a maximum value. Below this threshold, every intensity and phase state can be achieved. Even if this threshold is a small fraction of the maximum output intensity, it is not a limitation if the input intensity is high enough. To find such configurations, the concave hulls that contain the sets of states generated by sweeping over the gray levels were found. If a hull contains the origin of the complex plane, the closest point to the origin defines the radius of the largest centered circle that can be located inside of it. Then, limiting the intensity to such a maximum value, every complex amplitude state can be reached. Finally, the optimal configuration was considered to be the one with the largest radius.

\section{Results} \label{sec:res}
In order to test our optimization method, different configurations were set in the laboratory and characterized for a comparison with the numerical results. Firstly, the Mueller matrix of every optical element was experimentally determined to account for deviations from their theoretical counterparts. In this work, we used two Epson P09SG110 transmissive twisted nematic LCDs extracted from a Mitsubishi \mbox{LVP-SA51U} commercial video projector. Their dimensions are \mbox{18.5$\times$13.9 mm}, with a resolution of \mbox{804$\times$604} pixels, and a pixel pitch of approximately \mbox{20 $\mu$m}. By means of a diaphragm, we illuminated a region of \mbox{$\sim$ 200$\times$100} pixels in the central region of the displays. The region was selected in order to obtain an acceptable resolution for displaying images and avoiding border effects. As LCDs can be spatially non-uniform \cite{tiwari2021}, the characterization in terms of the gray level has to be performed specifically for the selected region.\\

In our experimental implementations the LCDs are separated by a few millimeters, with their holders in contact. This is the simplest double-pass configuration that allows for acceptable modulations and results in negligible diffraction effects. For other experimental setups, a PSC may be needed, e. g., a reflective mirror in the case in which two different zones of a reflective screen are used. Despite such a case requires pixel-to-pixel alignment, this extra experimental complexity with respect to the use of a single screen, still has some advantages. In fact, although there are SLMs, based on a different technology, that are capable of complex modulation, twisted nematic LCDs are widely used for commercial applications since they are inexpensive.\\

\begin{figure}[h]
\centering\includegraphics[width=0.9\linewidth]{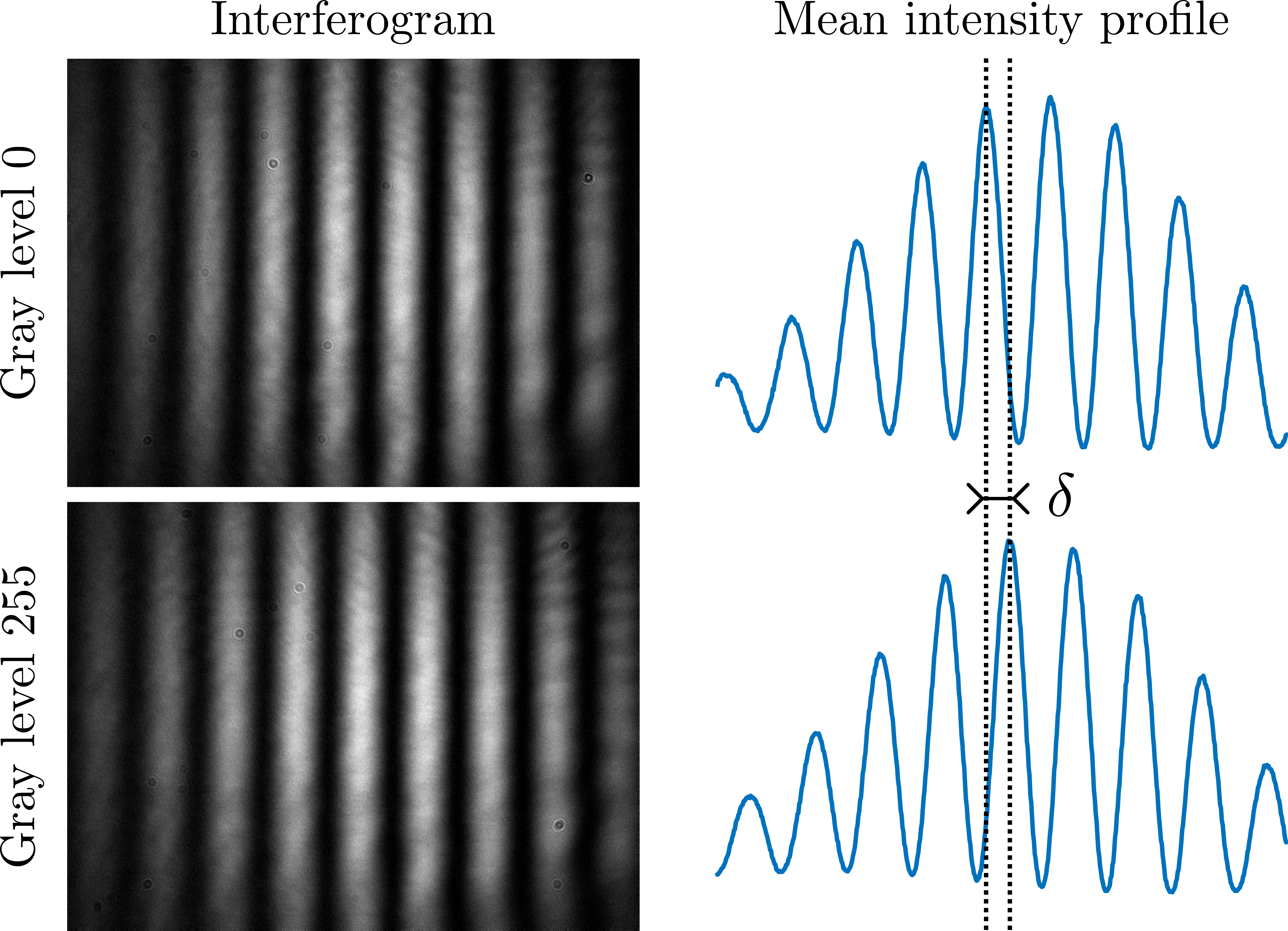}
\caption{Examples of the experimentally captured interferograms and their corresponding mean intensity profiles. From them, the displacement of the fringes $\delta$ is obtained, which allows to calculate the phase shift between the zero gray level and any other gray level (255 in this example).}
\label{fig:interf}
\end{figure}

The characterization of Mueller matrices was performed through a Mueller polarimeter, while global phase delays, $\beta_1$ and $\beta_2$ in Eq.~(\ref{eq:jonesvector}), were measured by interferometry. The incoming beam was splitted into two non-overlapping beams, each impinging onto a different half of the illuminated region of the LCDs. One of these regions is configured to display the gray level whose phase is being measured, while the other region, used as a reference, displays the zero gray level. Then the beams were collected with a lens to make them interfere. The resulting interferogram was magnified by a microscope objective and captured with a TheImagingSource \mbox{DMK 31BU03.H} camera. Finally, following a similar procedure as the one described in Ref.~\cite{moreno2003}, the phases were obtained from interferograms like the ones shown in Fig. \ref{fig:interf}. In this figure we show how different gray levels produce different fringe displacements, which is used to determine the phase shift relative to the reference level. It is then compared to the phase predicted by Jones calculus for that experimental configuration: the difference between the experimental phase and the predicted one is the global phase $\beta_i$. This procedure was repeated for several experimental configurations, from which the averaged $\beta_i$ values were obtained with a dispersion of around 5$^o$.\\

The gray levels of the displays in our setup can be independently set to 256 different values. However, to have a good characterization of its modulation properties, it was enough to measure in steps of 5 gray levels, giving a total of 52 measurements, throughout the dynamic range, for each single screen. In addition, to jointly characterize the double-pass setup, the sweep was performed over both gray levels simultaneously, being acquired a total of 52$^2$=2704 output states for each optical configuration. Due to the large number of data to be processed, we first ran coarse simulations, for which the orientations of the optical axes of the optical elements were considered in steps of 10$^o$ or 15$^o$. Then, finer simulations with steps of 5$^o$ were performed around the best results. Once the configurations that optimized the desired modulations were selected, the obtained simulated states according to the corresponding gray levels were compared to the experimentally obtained ones. These experimental measurements were performed sweeping over 18 gray levels in steps of 15, in each screen, which results in 18$^2$=324 experimental output states.\\

As we will show in this section, all the experimental results are in very good agreement with the corresponding simulations. Small differences between them can be explained as a consequence of small variations on the orientation of any of the optical elements during the characterizations or during the measurements themselves. Also, there may be fluctuations on the laser intensity or the electrical voltage applied to the displays by the used commercial video projector.

\subsection{Polarization modulation} \label{sec:res_pol}

\begin{figure}[h]
\centering\includegraphics[width=1\linewidth]{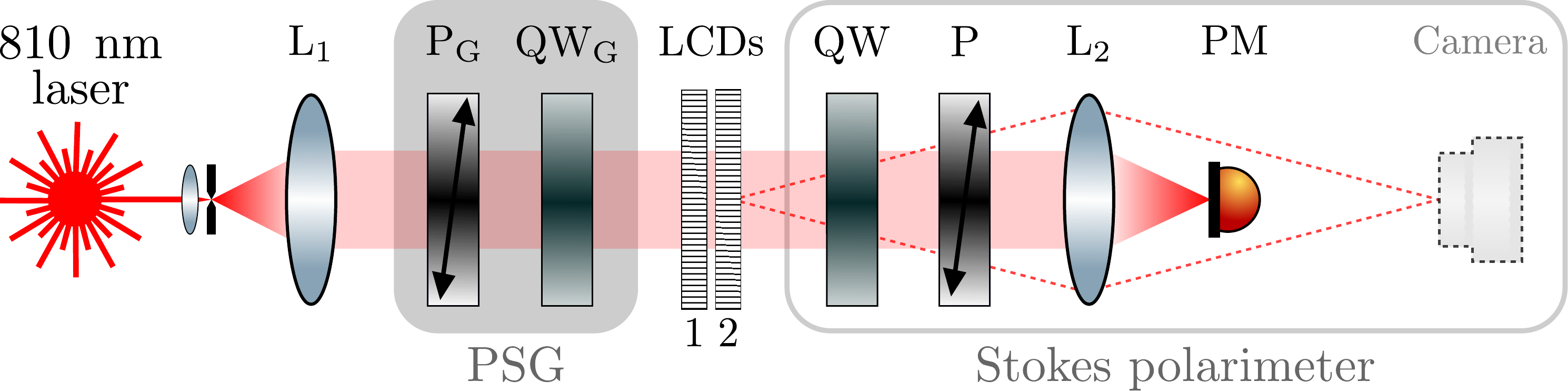}
\caption{Schematic of the experimental setup used for polarization modulation. An 810 nm laser beam is filtered, expanded and collimated. The generator system PSG comprises a polarizer (P$_\text{G}$) and a quarter waveplate (QW$_\text{G}$). Liquid crystal displays LCD$_i$ modulate the initial polarization states, which are then measured with a Stokes polarimeter that comprises a quarter waveplate QW, a polarizer P, and a power meter PM in the focal plane of a lens L$_2$. Alternatively, a Camera in the image plane of the displays can be used.}
\label{fig:pol1}
\end{figure}

The chosen system is the one shown in Fig. \ref{fig:pol1}, which is mathematically described by 

\begin{equation}
S_{\text{OUT}}=M_{\text{LCD}_2}\,  M_{\text{LCD}_1}\, M_{\text{QW}_\text{G}}\, M_{\text{P}_\text{G}}\, S_{\text{IN}}.
\label{eq:pol1}
\end{equation}

In this setup, a spatially filtered beam from a 810 nm laser source is expanded and collimated by lens L$_1$ of focal length 32 mm. In order to generate any input polarization state, the PSG includes a linear polarizer P$_\text{G}$ (measured extinction ratio \mbox{$\sim 1/4000$}) and a multi-order quarter waveplate QW$_\text{G}$, whose optical axes are set to angles $\theta_\text{PG}$ and $\theta_\text{QWG}$, respectively. The spatial modulation devices are the aforementioned displays, LCD$_1$ and LCD$_2$. In this system, neither the coupler (PSC) between the LCDs nor the analyzer (PSA) shown in Fig.~\ref{fig:setup_generico}, are present. Finally, the created polarization image $S_{\text{OUT}}$ was captured with a Stokes polarimeter, which in our case comprises a multi-order quarter waveplate QW, a Glan-Thompson polarizer P, and a Newport \mbox{1918-R} power meter located in the focal plane of a lens L$_2$ (focal length 200 mm) to measure the mean intensity of the illuminated region on the LCDs. Alternatively, a camera can be placed in the image plane of the LCDs to record the image formed by L$_2$.\\

\begin{figure}[h]
\centering\includegraphics[width=1\linewidth]{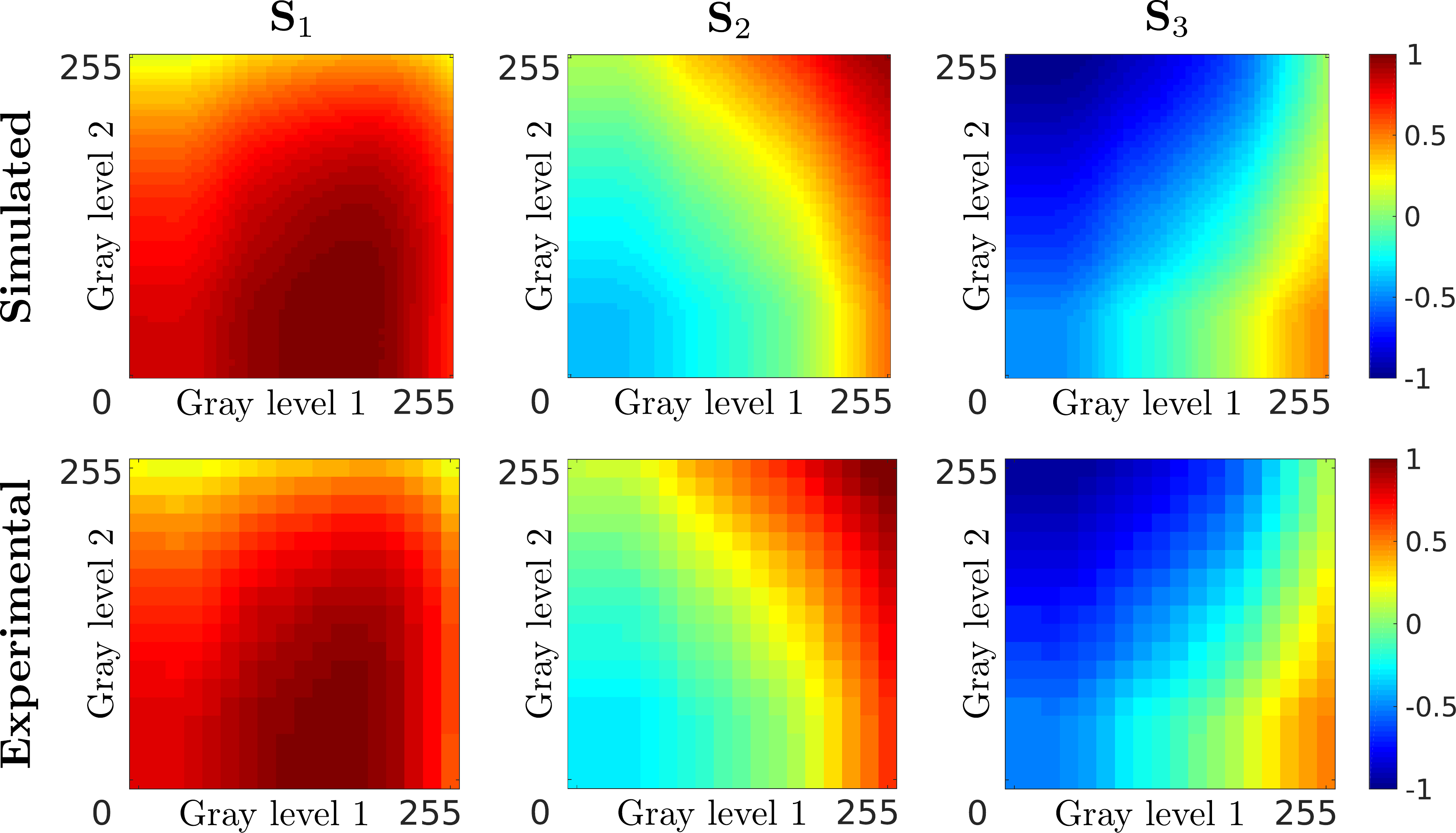}
\caption{Normalized Stokes parameters as a function of gray levels, for the setup shown in Fig. \ref{fig:pol1} with $\theta_{\text{PG}}=0^o$, $\theta_{\text{QWG}}=65^o$, optimized for maximum polarization modulation.}
\label{fig:stokes_0_65}
\end{figure}

\begin{figure}[h]
\centering\includegraphics[width=0.9\linewidth]{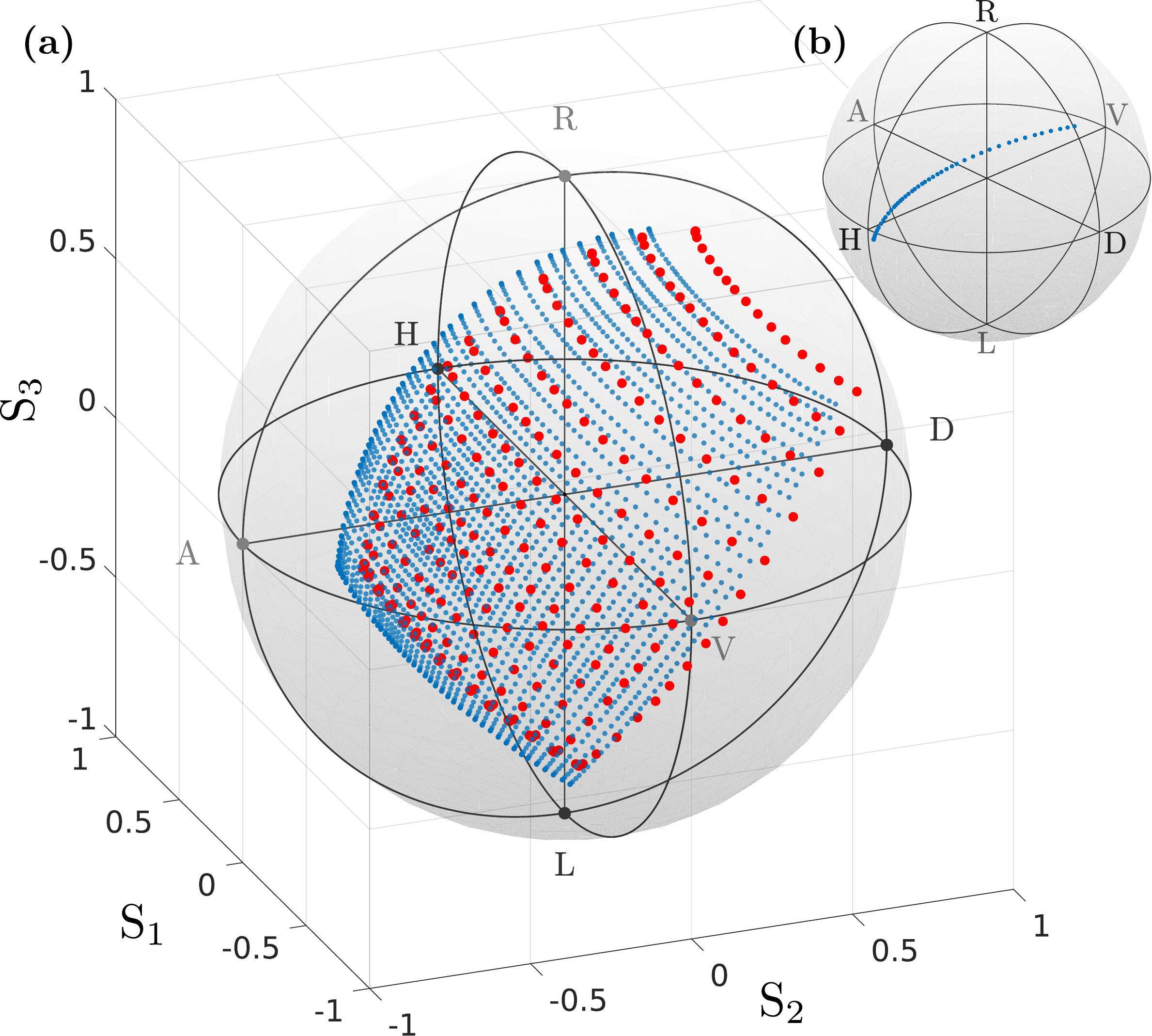}
\caption{Polarization results on the Poincaré sphere, for the setup shown in Fig. \ref{fig:pol1} with $\theta_{\text{PG}}=0^o$, $\theta_{\text{QWG}}=65^o$. (a) In blue the simulated states using experimental Mueller matrices, optimized for maximum modulation; in red the experimentally obtained ones. (b) Comparison with results for a single screen.}
\label{fig:esfera_0_65}
\end{figure}

After running the simulations, we have observed that a complete polarization modulation is not achievable with this setup, but we obtained that the best configuration is the one corresponding to $\theta_{\text{PG}}=0^o$ and $\theta_{\text{QWG}}=65^o$. In Fig. \ref{fig:stokes_0_65}, we show the color maps corresponding to the normalized Stokes parameters as a function of the gray levels of LCD$_1$ and LCD$_2$: in the first row the simulation results, and in the second row the experimentally obtained ones. In Fig. \ref{fig:esfera_0_65}(a) both results are represented in the Poincaré sphere. The degree of polarization is above $0.97$ for each of the considered states, while the mean Euclidean distance ($d$) to the four closest neighbor states is $0.020 \pm 0.005$. As a matter of comparison, these modulations were also attempted for a single-pass in one screen. For this case, the best results, shown in Fig. \ref{fig:esfera_0_65}(b), were obtained for $\theta_{\text{PG}}=0^o$ and $\theta_{\text{QWG}}=170^o$.\\

\begin{figure}[h]
\centering\includegraphics[width=1\linewidth]{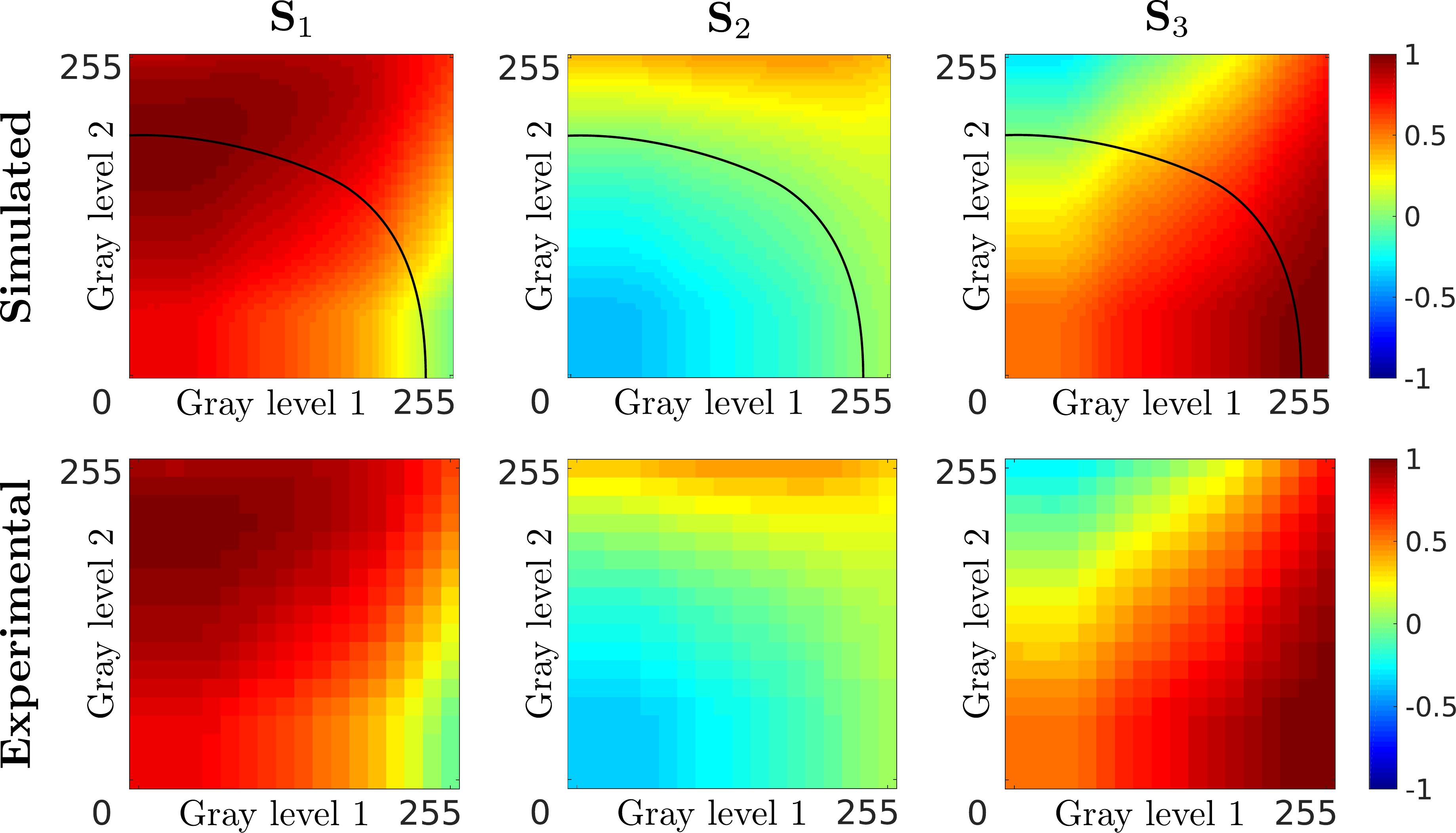}
\caption{Normalized Stokes parameters as a function of gray levels, for the setup shown in Fig. \ref{fig:pol1} with $\theta_{\text{PG}}=150^o$, $\theta_{\text{QWG}}=65^o$. In black, modulation along the RHCP-HLP meridian is shown.}
\label{fig:stokes_150_65}
\end{figure}

\begin{figure}[h]
\centering\includegraphics[width=0.9\linewidth]{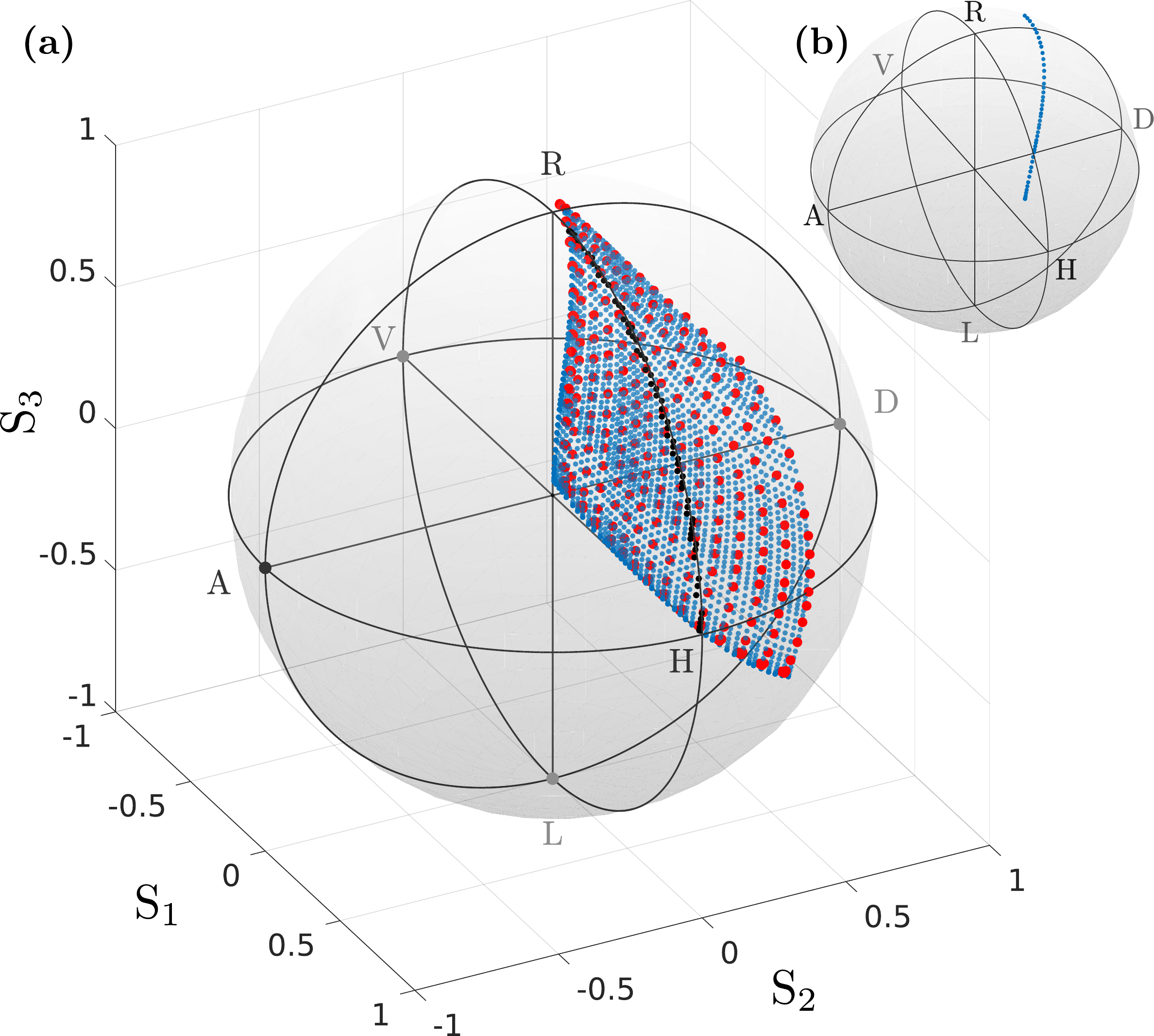}
\caption{Polarization results on the Poincaré sphere, for the setup shown in Fig. \ref{fig:pol1} with $\theta_{\text{PG}}=150^o$, $\theta_{\text{QWG}}=65^o$. (a) In blue the simulated states using experimental Mueller matrices, with the states along the meridian RHCP-HLP in black; in red the experimentally obtained ones. (b) Comparison with results for a single screen.}
\label{fig:esfera_150_65}
\end{figure}

We have also searched the configurations for particular polarization states of interest. As an example, for $\theta_{\text{PG}}=150^o$ and $\theta_{\text{QWG}}=65^o$, we reach states connecting right handed circular polarization (RHCP) with linear horizontal polarization (LHP) and laying near the meridian line. The normalized Stokes parameters, both the simulated and the experimentally obtained ones, are shown in Fig. \ref{fig:stokes_150_65}. A black line is added to visualize the combination of gray levels needed to achieve a modulation between RHCP and LHP that follows the meridian line. The corresponding Poincaré sphere is shown in Fig. \ref{fig:esfera_150_65}(a). Here, the obtained states that meet the requirement S$_2=0$, with a tolerance of $0.01$, are marked as black dots: they are a total of 68 distinguishable states with a degree of polarization above $0.98$. As a comparison, in Fig. \ref{fig:esfera_150_65}(b) it can be seen the states achievable with a single screen, for which the best results were obtained for $\theta_{\text{PG}}=165^o$ and $\theta_{\text{QWG}}=75^o$. In this case, only 17 states meet the condition S$_2=0$ with a tolerance of 0.10, and being the closest states to RHCP and HLP at an Euclidean distance of \mbox{$d\sim 0.41$}. 

\subsection{Complex amplitude modulation}

\begin{figure}[h]
\centering\includegraphics[width=1\linewidth]{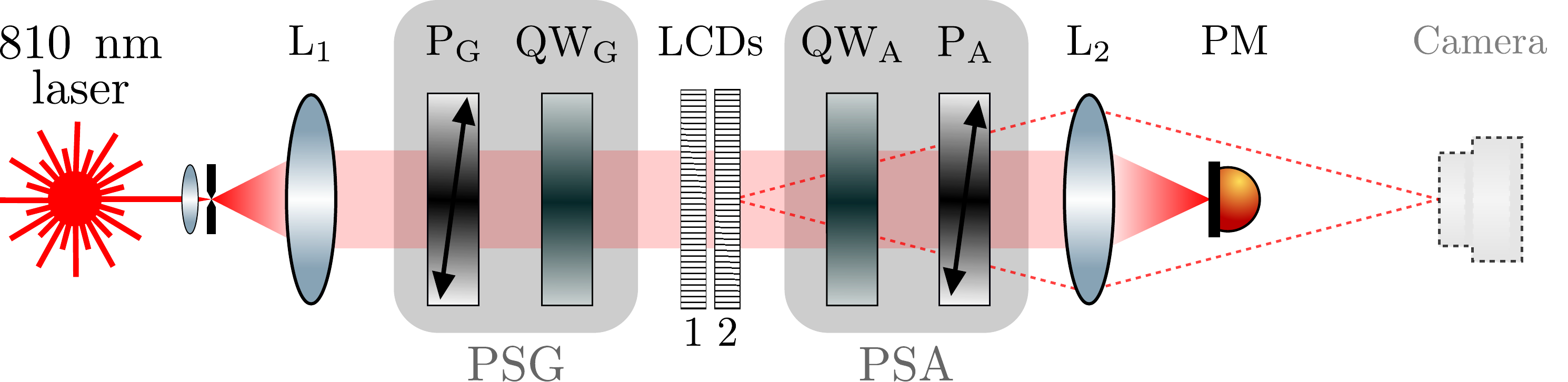}
\caption{Experimental setup used for a complex amplitude modulation. An 810 nm laser beam is filtered, expanded and collimated. The generator system PSG comprises a polarizer P$_\text{G}$ and a quarter waveplate QW$_\text{G}$. Liquid crystal displays LCD$_i$ modulate the initial polarization state, which is projected with the analyzer system PSA, that comprises a quarter waveplate QW$_\text{A}$ and a polarizer P$_\text{A}$. Then a power meter PM in the focal plane of a lens L$_2$ is used to measure intensity. Alternatively, a Camera in the image plane of the displays can be used.}
\label{fig:int1}
\end{figure}

The chosen system is the one shown in Fig. \ref{fig:int1}, which is mathematically described by

\begin{equation}
S_{\text{OUT}}=M_{\text{P}_\text{A}}\, M_{\text{QW}_\text{A}}\, M_{\text{LCD}_2} \, M_{\text{LCD}_1}\, M_{\text{QW}_\text{G}}\, M_{\text{P}_\text{G}}\, S_{\text{IN}}.
\label{eq:int1}
\end{equation}

It is worth noting that all the optical elements involved here are exactly the same as those shown in Fig. \ref{fig:pol1} and detailed in section \ref{sec:res_pol}, and configure the same experimental setup. However, they play a different role in the system given that, in that case, QW and P are not considered in the optimization but are part of the measurement stage. In the present case, they are part of the analyzer (PSA) and are therefore considered in the optimization process. Thus, the polarization states after the LCDs are projected by the PSA, which comprises a quarter waveplate QW$_\text{A}$ and a linear polarizer P$_\text{A}$, whose optical axes are set to angles $\theta_\text{QWA}$ and $\theta_\text{PA}$. Finally, a camera in the image plane of the lens L$_2$ can be used to capture the created intensity images $S_{\text{OUT}}$. Alternatively, we have used a power meter as described in section \ref{sec:res_pol} to measure the mean intensity of the illuminated region on the LCDs. As seen in Fig. \ref{fig:interf}, phases were measured from the interference between the generated states and a reference state, that one obtained when the gray levels of both LCD$_1$ and LCD$_2$ are set to zero.\\

\begin{figure}[h]
\centering\includegraphics[width=0.9\linewidth]{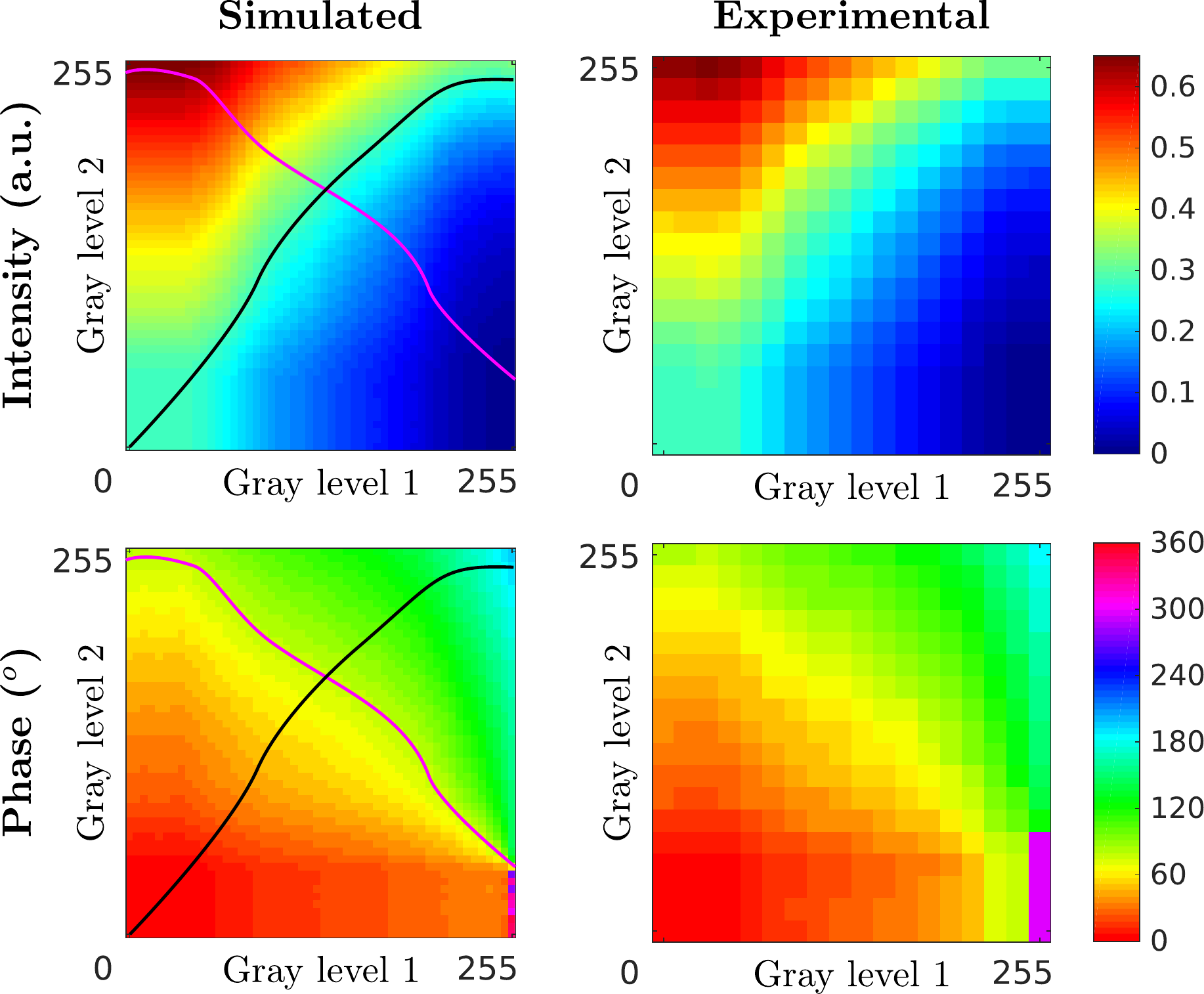}
\caption{Intensity and phase as a function of gray levels, for the setup in Fig. \ref{fig:int1} with $\theta_{\text{PG}}=75^o$, $\theta_{\text{QWG}}=105^o$, $\theta_{\text{QWA}}=15^o$, $\theta_{\text{PA}}=120^o$. In black and magenta, respectively, phase-only and intensity-only modulations are shown.}
\label{fig:calor_75_105_15_120}
\end{figure}

\begin{figure}[h]
\centering\includegraphics[width=1\linewidth]{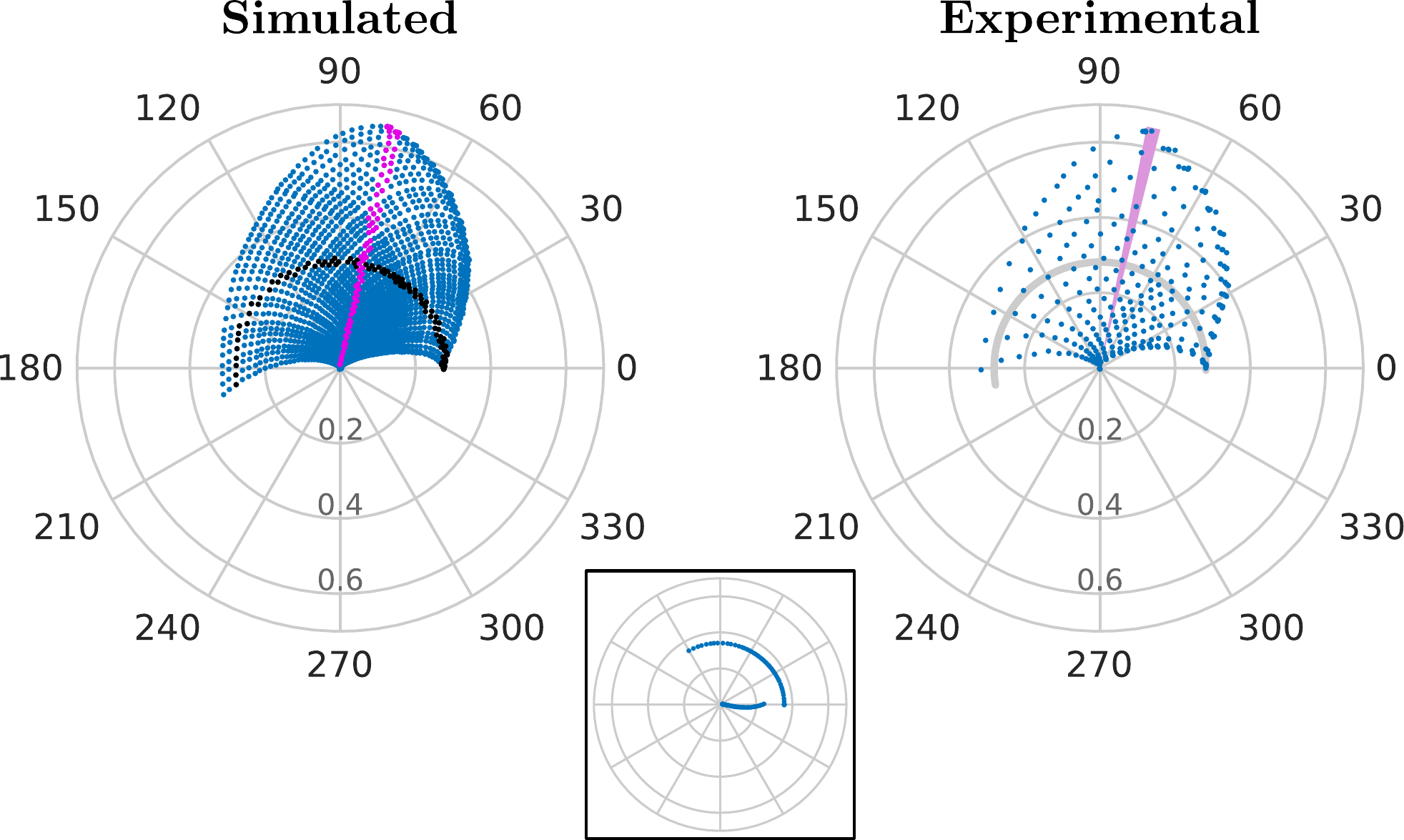}
\caption{Complex amplitude results in polar coordinates, for the setup shown in Fig. \ref{fig:int1} with $\theta_{\text{PG}}=75^o$, $\theta_{\text{QWG}}=105^o$, $\theta_{\text{QWA}}=15^o$, $\theta_{\text{PA}}=120^o$. In black and magenta, respectively, phase-only and intensity-only modulations are shown. In the inset, a comparison with optimizations for mostly phase-only and mostly intensity-only modulations using a single screen, each for a different experimental configuration.}
\label{fig:puntos_75_105_15_120}
\end{figure}

After running the simulation, we observed that a complete complex amplitude modulation is not achievable for these screens. Then we obtained the best configuration that simultaneously included an intensity-only and a phase-only modulation. This is the one with the setting parameters $\theta_{\text{PG}}=75^o$, $\theta_{\text{QWG}}=105^o$, $\theta_{\text{QWA}}=15^o$ and $\theta_{\text{PA}}=120^o$. In Fig. \ref{fig:calor_75_105_15_120}, the color maps show the intensity and phase for this configuration, as a function of the gray levels of the displays. The results obtained from the simulation (left column) are in very good agreement with those obtained from the experimental measurement (right column). Magenta lines are added to the simulated maps to visualize the combination of gray levels that allow to achieve intensity-only modulation while the phase remains constant. The obtained intensity range goes from $0.7\%$ to $65.1\%$ of the maximum intensity of the outcoming beam, resulting in a contrast of \mbox{$\sim 0.98$}. Also, the black lines show the combination of gray levels for which a phase-only modulation of $\pi$ radians is achieved. In Fig. \ref{fig:puntos_75_105_15_120}, we have represented each achievable state as a dot in a polar complex plane. Here, the length of the vector from the origin to a given point represents the intensity of the light state while the angle of such vector represents the phase. Magenta dots correspond to the 78 obtained states for intensity-only modulation, with phase $77^o\pm2^o$, while black dots correspond to the 30 states for phase-only modulation, with intensity $0.28\pm0.01$. For a single-pass setup only intensity or phase modulation can be achieved and, in consequence, it is not possible to cover two-dimensional surfaces on the complex plane, i.e., for a given optical configuration the generated states lie on curves. As a comparison, optimal configurations for intensity-only modulation and phase-only modulation were searched for one screen in a single-pass setup. For intensity-only modulation, the criterion was to maximize the contrast while keeping the phase variations below $10^o$. One of the configurations that meets this criterion is that corresponding to $\theta_{\text{PG}}=30^o$, $\theta_{\text{QWG}}=80^o$, $\theta_{\text{QWA}}=20^o$ and $\theta_{\text{PA}}=140^o$, with which 52 states can be generated whose intensities range from $1.5\%$ to $24.5\%$ (contrast \mbox{$\sim 0.88$}) and their phases are $-5^o \pm 5^o$. For phase-only modulation, we maximize the range in which the phase can be varied while keeping the intensity variations below 10\%. Fulfilling this criterion, the configuration with $\theta_{\text{PG}}=85^o$, $\theta_{\text{QWG}}=65^o$, $\theta_{\text{QWA}}=30^o$ and $\theta_{\text{PA}}=45^o$ allows to generate 52 states with intensity $0.35\pm0.02$ and spanning 2$\pi$/3 radians. The achievable states with these two experimental configurations can be seen in the inset of Fig. \ref{fig:puntos_75_105_15_120}.\\

\begin{figure}[h]
\centering\includegraphics[width=0.9\linewidth]{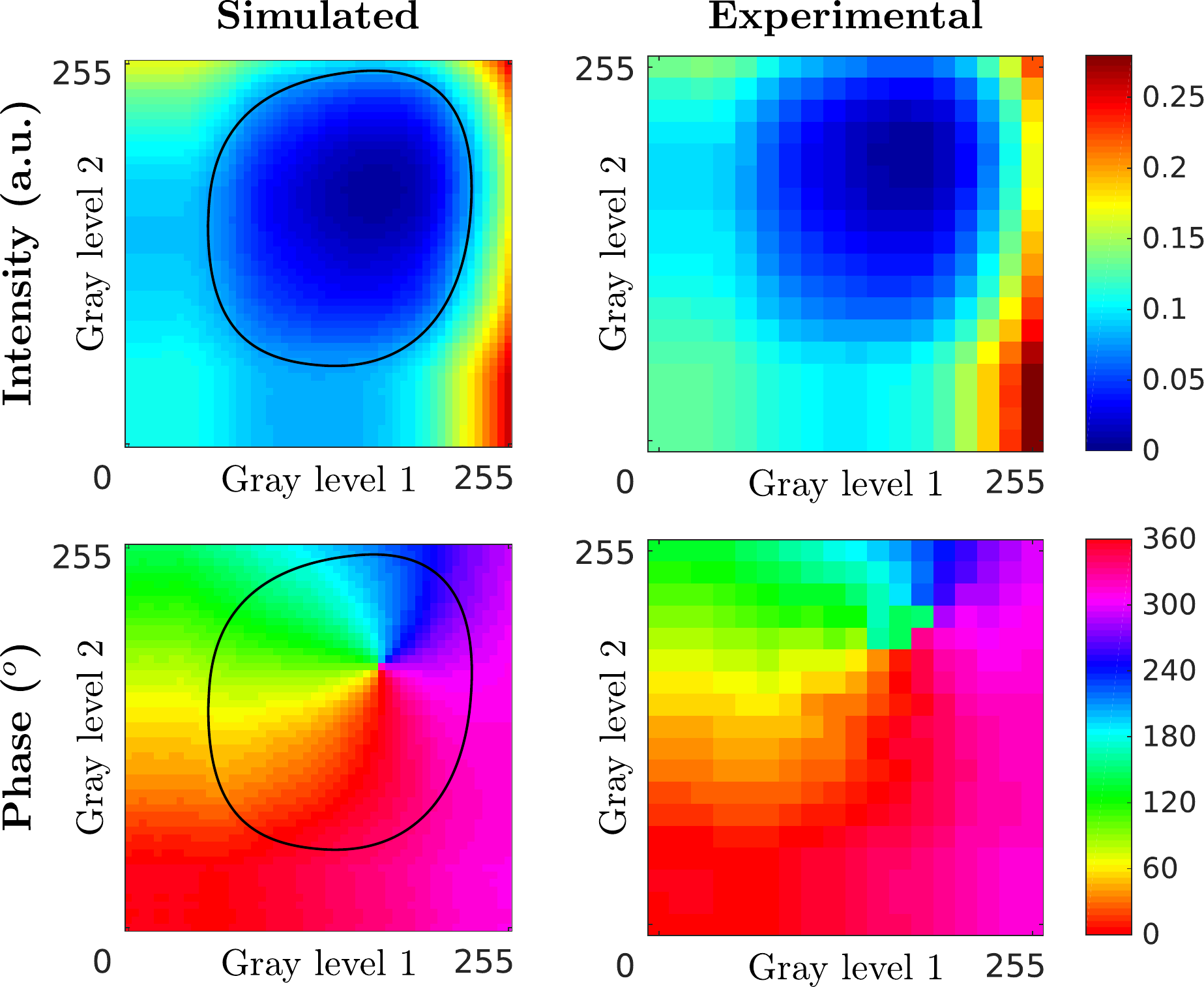}
\caption{Intensity and phase as a function of gray levels, for the setup in Fig. \ref{fig:int1} with $\theta_{\text{PG}}=-10^o$, $\theta_{\text{QWG}}=60^o$, $\theta_{\text{QWA}}=95^o$, $\theta_{\text{PA}}=85^o$. In black, the threshold for arbitrary complex amplitude modulation is shown.}
\label{fig:calor_350_60_95_85}
\end{figure}

\begin{figure}[h]
\centering\includegraphics[width=1\linewidth]{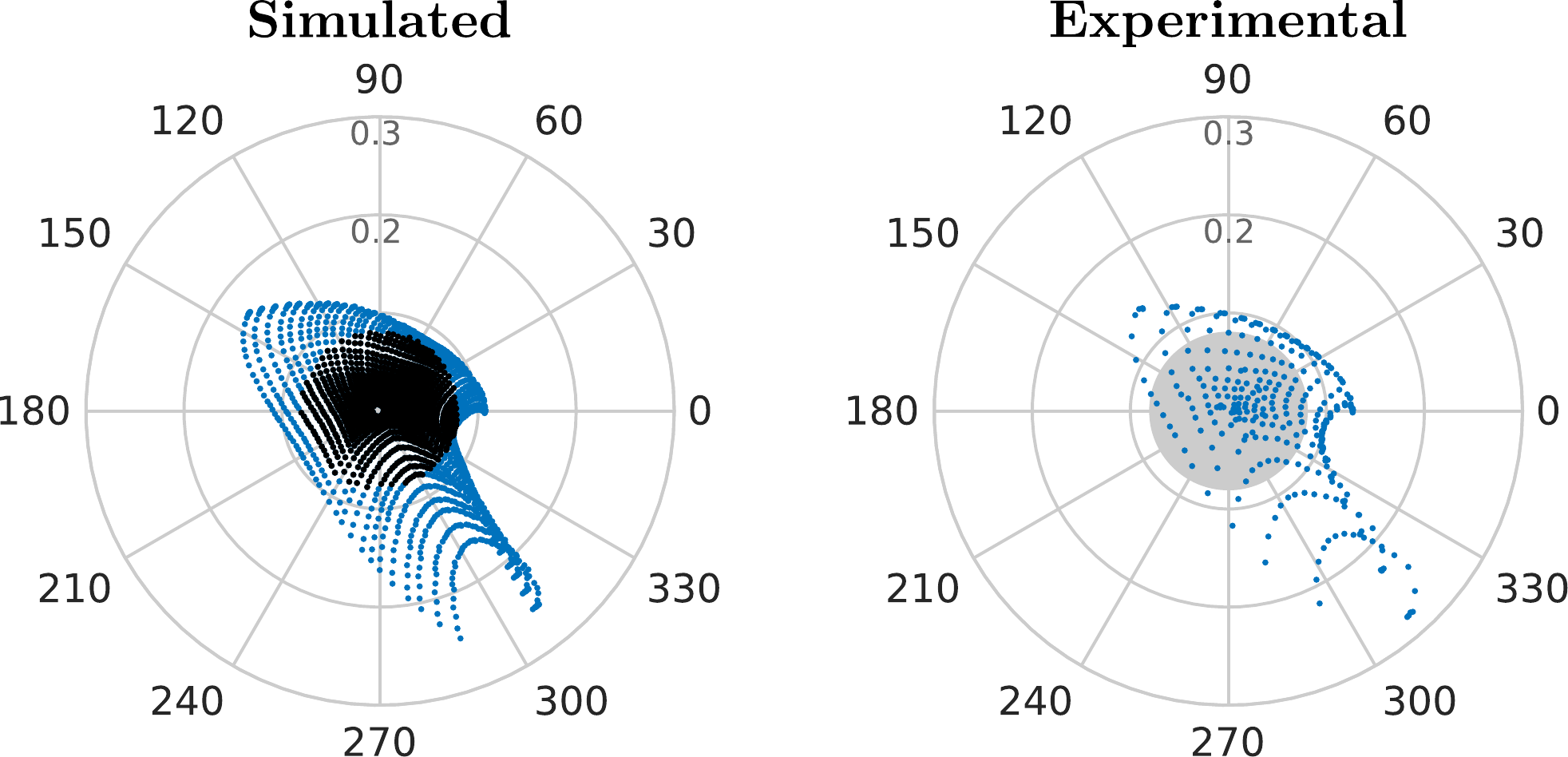}
\caption{Complex amplitude results in polar coordinates, for the setup shown in Fig. \ref{fig:int1} with $\theta_{\text{PG}}=-10^o$, $\theta_{\text{QWG}}=60^o$, $\theta_{\text{QWA}}=95^o$, $\theta_{\text{PA}}=85^o$. In black, the states for arbitrary complex amplitude modulation below an intensity threshold of $8\%$.}
\label{fig:puntos_350_60_95_85}
\end{figure}

Finally, as an example of arbitrary modulation below an intensity value, for $\theta_{\text{PG}}=-10^o$, $\theta_{\text{QWG}}=60^o$, $\theta_{\text{QWA}}=95^o$ and $\theta_{\text{PA}}=85^o$, we get a collection of states that allows us to reach every complex amplitude value if the intensity is limited to $8.0\%$ of the maximum output intensity, being the contrast \mbox{$\sim 0.86$}. It should be pointed out that this threshold is not a limitation if the intensity of the incoming field is high enough: for example, in our case it means an optical power of \mbox{$60 \mu$W} for a laser of \mbox{$\sim 50$ mW}. In Fig. \ref{fig:calor_350_60_95_85}, that threshold is shown as a black line. This corresponds, in Fig. \ref{fig:puntos_350_60_95_85}, to 1246 states with intensities lower than $0.080$ (black dots). Such a result can not be obtained with a single-pass setup, given that the generated states always lay on a curve in the complex plane, in contrast with the 2D surfaces obtained by modulating, as in our case, with two screens.

\section{Conclusions}
In this work, we have implemented an optimization method for optical setups that include a double-pass SLM system. The Mueller matrices of the SLMs and  passive linear optic elements, as well as the phase delays they introduce, are taken as inputs of the optimization algorithm. This searches for the optimal optical configuration that maximizes the modulation capabilities of the whole system. To fully explore these capabilities, for any desired modulation, the search was done by considering the total effect of both impingements, that is, the optimization is carried out on the two SLMs working as a single device with more free parameters, and not as two independent devices working together. By doing this, not only the modulation performance is improved in comparison with an only screen in a single-pass setup, but other modulations of the electromagnetic field properties are achieved. In fact, simultaneously changing the gray level of both screens (or of the two zones of the same screen onto which the light impinges twice) generates states of light that cover a two-dimensional surface, either on the complex plane or on the Poincaré sphere, while a single screen can only generate curves. Although it is possible to obtain two-dimensional surfaces with a single-pass setup by encoding one degree of freedom into another one or combining several individual pixels into a macro-pixel, this requires more complex experimental setups and does not allow to use the full resolution of the display.\\

Our optimization method has been used to conduct numerical simulations for different modulations of the electromagnetic field properties, and the results were compared with their experimental counterparts, showing a very good agreement within the numerical and experimental errors. In particular, we tested the method for a setup involving two twisted nematic LCDs. Two modulations were studied: polarization and complex amplitude modulation. To our knowledge, the first one has not been studied so far in the case of double-pass configurations using this approach. The second one has been treated in \cite{macfaden2017} following the Jones formalism, which does not allow to take into account unwanted depolarization effects that usually have to be dealt with in real experimental situations.\\

As highlights we can mention that the use of this optimized double-pass setup squares the amount of generated states, in our case going from 256 for a single-pass setup, up to 65536. When studying polarization modulation we observed that not only the dynamical range is incremented with respect to a single-pass setup, but also there is an order of magnitude improvement in both the resolution and the accuracy with which the desired states can be obtained. In the case of complex amplitude modulation, the optimization allowed us to find the configuration for a complete modulation in the complex plane below a maximum intensity threshold, and also that which makes it possible to modulate pure phase and pure intensity without modifying the optical configuration. None of them can be achieved using a setup in a single-pass configuration.\\

Finally, we want to emphasize that although the results of each optimization are strongly dependent on the chosen SLMs, the presented approach provides a simple method to make the most out of any SLM and it is versatile enough to deal with different experimental setups.

\begin{backmatter}
\bmsection{Funding}
This work was supported by CONICET-PIP 112200801-03047 and UBACyT 20020170100564BA.
\bmsection{Acknowledgments}
Sebastián Bordakevich was supported by a CONICET fellowship.
\bmsection{Disclosures}
The authors declare no conflicts of interest.
\bmsection{Data availability} Data underlying the results presented in this paper are not publicly available at this time but may be obtained from the authors upon reasonable request.
\end{backmatter}

\bibliography{main}

\begin{thebibliography}{10}
\newcommand{\enquote}[1]{``#1''}

\bibitem{maurer2010}
C.~Maurer, A.~Jesacher, S.~Bernet, and M.~Ritsch-Marte, \enquote{What spatial
  light modulators can do for optical microscopy,} {\protect\JournalTitle{Laser
  \& Photonics Reviews}} \textbf{5}, 81--101 (2011).

\bibitem{trichili2016}
A.~Trichili, C.~Rosales-Guzm{\'a}n, A.~Dudley, B.~Ndagano, A.~Ben~Salem,
  M.~Zghal, and A.~Forbes, \enquote{Optical communication beyond orbital
  angular momentum,} {\protect\JournalTitle{Scientific Reports}} \textbf{6},
  27674 (2016).

\bibitem{peinado2010}
A.~Peinado, A.~Lizana, J.~Vidal, C.~Iemmi, and J.~Campos, \enquote{Optimization
  and performance criteria of a stokes polarimeter based on two variable
  retarders,} {\protect\JournalTitle{Opt. Express}} \textbf{18}, 9815--9830
  (2010).

\bibitem{rubinsztein2017}
H.~Rubinsztein-Dunlop, A.~Forbes, M.~V. Berry, M.~R. Dennis, D.~L. Andrews,
  M.~Mansuripur, C.~Denz, C.~Alpmann, P.~Banzer, T.~Bauer, E.~Karimi,
  L.~Marrucci, M.~Padgett, M.~Ritsch-Marte, N.~M. Litchinitser, N.~P. Bigelow,
  C.~Rosales-Guzm{\'{a}}n, A.~Belmonte, J.~P. Torres, T.~W. Neely, M.~Baker,
  R.~Gordon, A.~B. Stilgoe, J.~Romero, A.~G. White, R.~Fickler, A.~E. Willner,
  G.~Xie, B.~McMorran, and A.~M. Weiner, \enquote{Roadmap on structured light,}
  {\protect\JournalTitle{Journal of Optics}} \textbf{19}, 013001 (2016).

\bibitem{bhebhe2018}
N.~Bhebhe, P.~A.~C. Williams, C.~Rosales-Guzm{\'a}n, V.~Rodriguez-Fajardo, and
  A.~Forbes, \enquote{A vector holographic optical trap,}
  {\protect\JournalTitle{Scientific Reports}} \textbf{8}, 17387 (2018).

\bibitem{flamini2018}
F.~Flamini, N.~Spagnolo, and F.~Sciarrino, \enquote{Photonic quantum
  information processing: a review,} {\protect\JournalTitle{Reports on Progress
  in Physics}} \textbf{82}, 016001 (2018).

\bibitem{pabon2019}
D.~Pab\'on, L.~Reb\'on, S.~Bordakevich, N.~Gigena, A.~Boette, C.~Iemmi,
  R.~Rossignoli, and S.~Ledesma, \enquote{Parallel-in-time optical simulation
  of history states,} {\protect\JournalTitle{Phys. Rev. A}} \textbf{99}, 062333
  (2019).

\bibitem{lohrmann2019}
A.~Lohrmann, C.~Perumgatt, and A.~Ling, \enquote{Manipulation and measurement
  of quantum states with liquid crystal devices,} {\protect\JournalTitle{Opt.
  Express}} \textbf{27}, 13765--13772 (2019).

\bibitem{chandrasekhar}
S.~Chandrasekhar, \emph{Liquid Crystals} (Cambridge University Press, UK,
  1992), 2nd ed.

\bibitem{neto1996}
L.~G. Neto, D.~Roberge, and Y.~Sheng, \enquote{Full-range, continuous, complex
  modulation by the use of two coupled-mode liquid-crystal televisions,}
  {\protect\JournalTitle{Appl. Opt.}} \textbf{35}, 4567--4576 (1996).

\bibitem{marquez2001}
A.~Marquez, C.~C. Iemmi, I.~S. Moreno, J.~A. Davis, J.~Campos, and M.~J. Yzuel,
  \enquote{{Quantitative prediction of the modulation behavior of twisted
  nematic liquid crystal displays based on a simple physical model},}
  {\protect\JournalTitle{Optical Engineering}} \textbf{40}, 2558 -- 2564
  (2001).

\bibitem{kruger2015}
M.~Kr\"{u}ger, R.~Kampmann, R.~Kleindienst, and S.~Sinzinger,
  \enquote{Time-resolved combination of the mueller-stokes and jones calculus
  for the optimization of a twisted-nematic spatial-light modulator,}
  {\protect\JournalTitle{Appl. Opt.}} \textbf{54}, 4239--4248 (2015).

\bibitem{chandra2020}
A.~D. Chandra and A.~Banerjee, \enquote{Rapid phase calibration of a spatial
  light modulator using novel phase masks and optimization of its efficiency
  using an iterative algorithm,} {\protect\JournalTitle{Journal of Modern
  Optics}} \textbf{67}, 628--637 (2020).

\bibitem{tiwari2021}
V.~Tiwari, Y.~Pandey, and N.~S. Bisht, \enquote{Spatially addressable
  polarimetric calibration of reflective-type spatial light modulator using
  mueller–stokes polarimetry,} {\protect\JournalTitle{Frontiers in Physics}}
  \textbf{9}, 400 (2021).

\bibitem{marquez2008}
A.~M\'{a}rquez, I.~Moreno, C.~Iemmi, A.~Lizana, J.~Campos, and M.~J. Yzuel,
  \enquote{Mueller-stokes characterization and optimization of a liquid crystal
  on silicon display showing depolarization,} {\protect\JournalTitle{Opt.
  Express}} \textbf{16}, 1669--1685 (2008).

\bibitem{moreno2008}
I.~Moreno, A.~Lizana, J.~Campos, A.~M\'{a}rquez, C.~Iemmi, and M.~J. Yzuel,
  \enquote{Combined mueller and jones matrix method for the evaluation of the
  complex modulation in a liquid-crystal-on-silicon display,}
  {\protect\JournalTitle{Opt. Lett.}} \textbf{33}, 627--629 (2008).

\bibitem{kelly1999}
T.~Kelly and J.~Munch, \enquote{Genetic optimization of modulation
  characteristics for two twisted nematic liquid crystal spatial light
  modulators,} {\protect\JournalTitle{Optical and Quantum Electronics}}
  \textbf{31}, 515--523 (1999).

\bibitem{hsieh2007}
M.-L. Hsieh, M.-L. Chen, and C.-J. Cheng, \enquote{{Improvement of the complex
  modulated characteristic of cascaded liquid crystal spatial light modulators
  by using a novel amplitude compensated technique},}
  {\protect\JournalTitle{Optical Engineering}} \textbf{46}, 1 -- 3 (2007).

\bibitem{lima2011}
G.~Lima, L.~Neves, R.~Guzm\'{a}n, E.~S. G\'{o}mez, W.~A.~T. Nogueira,
  A.~Delgado, A.~Vargas, and C.~Saavedra, \enquote{Experimental quantum
  tomography of photonic qudits via mutually unbiased basis,}
  {\protect\JournalTitle{Opt. Express}} \textbf{19}, 3542--3552 (2011).

\bibitem{vanputten2008}
E.~G. van Putten, I.~M. Vellekoop, and A.~P. Mosk, \enquote{Spatial amplitude
  and phase modulation using commercial twisted nematic lcds,}
  {\protect\JournalTitle{Appl. Opt.}} \textbf{47}, 2076--2081 (2008).

\bibitem{maluenda2013}
D.~Maluenda, I.~Juvells, R.~Mart\'{i}nez-Herrero, and A.~Carnicer,
  \enquote{Reconfigurable beams with arbitrary polarization and shape
  distributions at a given plane,} {\protect\JournalTitle{Opt. Express}}
  \textbf{21}, 5432--5439 (2013).

\bibitem{hasegawa2019}
S.~ya~Hasegawa and H.~Inoue, \enquote{High spatial resolution pixel synthesis
  structure for full-complex amplitude modulation with twisted nematic lcd,}
  {\protect\JournalTitle{Appl. Opt.}} \textbf{58}, 6725--6732 (2019).

\bibitem{davis1999}
J.~A. Davis, D.~M. Cottrell, J.~Campos, M.~J. Yzuel, and I.~Moreno,
  \enquote{Encoding amplitude information onto phase-only filters,}
  {\protect\JournalTitle{Appl. Opt.}} \textbf{38}, 5004--5013 (1999).

\bibitem{arrizon2007}
V.~Arriz\'{o}n, U.~Ruiz, R.~Carrada, and L.~A. Gonz\'{a}lez, \enquote{Pixelated
  phase computer holograms for the accurate encoding of scalar complex fields,}
  {\protect\JournalTitle{J. Opt. Soc. Am. A}} \textbf{24}, 3500--3507 (2007).

\bibitem{bolduc2013}
E.~Bolduc, N.~Bent, E.~Santamato, E.~Karimi, and R.~W. Boyd, \enquote{Exact
  solution to simultaneous intensity and phase encryption with a single
  phase-only hologram,} {\protect\JournalTitle{Opt. Lett.}} \textbf{38},
  3546--3549 (2013).

\bibitem{varga2014}
J.~J.~M. Varga, L.~Reb{\'{o}}n, M.~A. Sol{\'{\i}}s-Prosser, L.~Neves,
  S.~Ledesma, and C.~Iemmi, \enquote{Optimized generation of spatial qudits by
  using a pure phase spatial light modulator,} {\protect\JournalTitle{Journal
  of Physics B: Atomic, Molecular and Optical Physics}} \textbf{47}, 225504
  (2014).

\bibitem{wang2007}
X.-L. Wang, J.~Ding, W.-J. Ni, C.-S. Guo, and H.-T. Wang, \enquote{Generation
  of arbitrary vector beams with a spatial light modulator and a common path
  interferometric arrangement,} {\protect\JournalTitle{Opt. Lett.}}
  \textbf{32}, 3549--3551 (2007).

\bibitem{rong2014}
Z.-Y. Rong, Y.-J. Han, S.-Z. Wang, and C.-S. Guo, \enquote{Generation of
  arbitrary vector beams with cascaded liquid crystal spatial light
  modulators,} {\protect\JournalTitle{Opt. Express}} \textbf{22}, 1636--1644
  (2014).

\bibitem{guo2014}
C.-S. Guo, Z.-Y. Rong, and S.-Z. Wang, \enquote{Double-channel vector spatial
  light modulator for generation of arbitrary complex vector beams,}
  {\protect\JournalTitle{Opt. Lett.}} \textbf{39}, 386--389 (2014).

\bibitem{macfaden2017}
A.~J. Macfaden and T.~D. Wilkinson, \enquote{Characterization, design, and
  optimization of a two-pass twisted nematic liquid crystal spatial light
  modulator system for arbitrary complex modulation,} {\protect\JournalTitle{J.
  Opt. Soc. Am. A}} \textbf{34}, 161--170 (2017).

\bibitem{juday1991}
R.~D. Juday and J.~M. Florence, \enquote{{Full-complex modulation with two
  one-parameter SLMs},} in \emph{Wave Propagation and Scattering in Varied
  Media II,}  vol. 1558 V.~K. Varadan, ed., International Society for Optics
  and Photonics (SPIE, 1991), pp. 499 -- 504.

\bibitem{sit2017}
A.~Sit, L.~Giner, E.~Karimi, and J.~S. Lundeen, \enquote{General lossless
  spatial polarization transformations,} {\protect\JournalTitle{Journal of
  Optics}} \textbf{19}, 094003 (2017).

\bibitem{kenny2012}
F.~Kenny, D.~Lara, O.~G. Rodr\'{i}guez-Herrera, and C.~Dainty,
  \enquote{Complete polarization and phase control for focus-shaping in high-na
  microscopy,} {\protect\JournalTitle{Opt. Express}} \textbf{20}, 14015--14029
  (2012).

\bibitem{goldstein}
D.~Goldstein, \emph{Polarized Light} (CRC Press, USA, 2003), 2nd ed.

\bibitem{boundary}
{MathWorks}, \enquote{{MATLAB official documentation},}
  \url{www.mathworks.com/help/matlab/ref/boundary.html}.

\bibitem{moreno2003}
I.~Moreno, P.~Velásquez, C.~R. Fernández-Pousa, M.~M. Sánchez-López, and
  F.~Mateos, \enquote{Jones matrix method for predicting and optimizing the
  optical modulation properties of a liquid-crystal display,}
  {\protect\JournalTitle{Journal of Applied Physics}} \textbf{94}, 3697--3702
  (2003).

\end{thebibliography}

\end{document}